# THE PIGEONHOLE BOOTSTRAP[1]

By Art B. Owen

*Stanford University*

Recently there has been much interest in data that, in statistical language, may be described as having a large crossed and severely unbalanced random effects structure. Such data sets arise for recommender engines and information retrieval problems. Many large bipartite weighted graphs have this structure too. We would like to assess the stability of algorithms fit to such data. Even for linear statistics, a naive form of bootstrap sampling can be seriously misleading and McCullagh [*Bernoulli* **6** (2000) 285–301] has shown that no bootstrap method is exact. We show that an alternative bootstrap separately resampling rows and columns of the data matrix satisfies a mean consistency property even in heteroscedastic crossed unbalanced random effects models. This alternative does not require the user to fit a crossed random effects model to the data.

**1. Introduction.** Many important statistical problems feature two interlocking sets of entities, customarily arranged as rows and columns. Unlike the usual cases by variables layout, these data fit better into a cases by cases interpretation. Examples include books and customers for a web site, movies and raters for a recommender engine, and terms and documents in information retrieval. Historically data with this structure has been studied with a crossed random effects model. The new data sets are very large and haphazardly structured, a far cry from the setting for which normal theory random effects models were developed. It can be hard to estimate the variance of features fit to data of this kind.

Parametric likelihood and Bayesian methods typically come with their own internally valid methods of estimating variances. However, the crossed random effects setting can be more complicated than what our models anticipate. If in IID sampling we suspect that our model is inadequate, then we can make a simple and direct check on it via bootstrap resampling of

---

Received May 2007; revised May 2007.
[1]Supported by NSF Grant DMS-06-04939.

cases. We can even judge sampling uncertainty for computations that were not derived from any explicit model.

We would like to have a version of the bootstrap suitable to large unbalanced crossed random effect data sets. Unfortunately for those hopes, McCullagh (2000) has proved that no such bootstrap can exist, even for the basic problem of finding the variance of the grand mean of the data in a balanced setting with no missing values and homoscedastic variables.

McCullagh (2000) included two reasonably well performing approximate methods for balanced data sets. They yielded a variance that was nearly correct under reasonable assumptions about the problem. One approach was to fit the random effects model and then resample from it. That option is not attractive for the kind of data set considered here. Even an oversimplified model can be hard to fit to unbalanced data, and the results will lack the face value validity that we get from the bootstrap for the IID case. The second method resampled rows and columns independently. This approach imitates the Cornfield and Tukey (1956) pigeonhole sampling model, and is preferable operationally. We call it the pigeonhole bootstrap, and show that it continues to be a reasonable estimator of variance even for seriously unbalanced data sets and inhomogenous (nonexchangeable) random effects models.

In notation to be explained further below, we find that the true variance of our statistic takes the form $(\nu_A \sigma_A^2 + \nu_B \sigma_B^2 + \sigma_E^2)/N$, where $\nu_A$ and $\nu_B$ can be calculated from the data and satisfy $1 \ll \nu \ll N$ in our motivating applications. A naive bootstrap (resampling cases) will produce a variance estimate close to $(\sigma_A^2 + \sigma_B^2 + \sigma_E^2)/N$ and thus be seriously misleading. The pigeonhole bootstrap will produce a variance estimate close to $((\nu_A + 2)\sigma_A^2 + (\nu_B + 2)\sigma_B^2 + 3\sigma_E^2)/N$. It is thus mildly conservative, but not unduly so in cases where each $\nu \gg 2$ and $\sigma_E^2$ does not dominate.

McCullagh (2000) leaves open the possibility that a linear combination of several bootstrap methods will be suitable. In the present setting the pigeonhole bootstrap overestimates the variance by twice the amount of the naive bootstrap. One could therefore bootstrap both ways and subtract twice the naive variance from the pigeonhole variance. That approach, of course, brings the usual difficulties of possibly negative variance estimates. Also, sometimes we do not want the variance per se, just a histogram that we think has approximately the right width, and the variance is only a convenient way to decide if a histogram has roughly the right width. Simply accepting a bootstrap histogram that is slightly too wide may be preferable to trying to make it narrower by an amount based on the naive variance.

Many of the motivating problems come from e-commerce. There one may have to decide where on a web page to place an ad or which book to recommend. Because the data sets are so large, coarse granularity statistics can be estimated with essentially negligible sampling uncertainty. For example, the Netflix data set has over 100 million movie ratings and the average movie rating is very well determined. Finer, subtle points, such as whether classical music lovers are more likely to purchase a Harry Potter book on a Tuesday are a different matter. Some of these may be well determined and some will not. An e-commerce application can keep track of millions of subtle rules, and the small advantages so obtained can add up to something commercially valuable. Thus, the dividing line between noise artifacts and real signals is worth finding, even in problems with large data sets.

The outline of this paper is as follows. Section 2 introduces the notation for row and column entities and sample sizes, including the critical quantities $\nu_A$ and $\nu_B$, as well as the random effects model we consider and the linear statistics we investigate. Section 3 introduces two bootstrap models: the naive bootstrap and the pigeonhole bootstrap. Section 4 derives the variance expressions we need. Section 5 presents a small example, using the bootstrap to determine whether movie ratings at Netflix that were made on a Tuesday really are lower than those from other days. There is a discussion in Section 6 including application to models using outer products as commonly fit by the SVD. The Appendix contains the proof of Theorem 3. Shorter proofs appear inline, but can be easily skipped over on first reading.

**2. Notation.** The row entities are $i = 1, \ldots, R$ and the column entities are $j = 1, \ldots, C$. The variable $Z_{ij} \in \{0, 1\}$ takes the value 1 if we have data for the $(i, j)$ combination and is 0 otherwise. The value $X_{ij} \in \mathbb{R}^d$ holds the observed data when $Z_{ij} = 1$ and otherwise it is missing. To hide inessential details we will work with $d = 1$, apart from a remark in Section 4.4.

The data are $X_{ij}$ for $i = 1, \ldots, R$ and $j = 1, \ldots, C$ for those $ij$ pairs with $Z_{ij} = 1$. The number of times that row $i$ was seen is $n_{i\bullet} = \sum_{j=1}^{C} Z_{ij}$. Similarly, column $j$ was seen $n_{\bullet j} = \sum_{i=1}^{R} Z_{ij}$ times. The total sample size is $N = n_{\bullet\bullet} = \sum_i \sum_j Z_{ij}$.

In addition to the $R \times C$ layout with missing entries described above, we can also arrange the data as a sparse matrix via an $N \times 3$ array $S$ with $\ell$th row $(I_\ell, J_\ell, X_\ell)$ for $\ell = 1, \ldots, N$. The value $X_\ell$ in this layout equals $X_{I_\ell J_\ell}$ from the $R \times C$ layout. The value $I_\ell = i$ appears $n_{i\bullet}$ times in column 1 of $S$, and similarly, $J_\ell = j$ appears $n_{\bullet j}$ times in column 2.

The ratios

$$\nu_A \equiv \frac{1}{N} \sum_{i=1}^{R} n_{i\bullet}^2 \quad \text{and} \quad \nu_B \equiv \frac{1}{N} \sum_{j=1}^{C} n_{\bullet j}^2,$$

prove to be important later. The value of $\nu_A$ is the expectation of $n_{i\bullet}$ when $i$ is sampled with probability proportional to $n_{i\bullet}$. If two not necessarily distinct observations having the same $i$ are called "row neighbors," then $\nu_A$ is the average number of row neighbors for observations in the data set. Similarly, $\nu_B$ is the average number of column neighbors.

In the extreme where no column has been seen twice, every $n_{\bullet j} = 1$ and then $\nu_B = 1$. In the other extreme where there is only one column, having $n_{\bullet 1} = N$, then $\nu_B = N$. Typically encountered problems should have $1 \ll \nu_B \ll N$ and $1 \ll \nu_A \ll N$. For example, an $R \times C$ table with no missing values has $\nu_B = C$ and we may expect both $R$ and $C$ to be large. Often there will be a Zipf-like distribution on $n_{\bullet j}$, and then for large problems we will find that $1 \ll \nu_B \ll N$. Similarly, cases with $1 \ll \nu_A \ll N$ are to be expected.

For the Netflix data set, $\nu$ for customers is about 646 and $\nu$ for movies is about 56,200. As we will see below, these large values mean that naive sampling models seriously underestimate the variance.

We also need the quantities

$$\mu_{\bullet j} \equiv \frac{1}{N} \sum_i Z_{ij} n_{i\bullet}$$

and

$$\mu_{i\bullet} \equiv \frac{1}{N} \sum_j Z_{ij} n_{\bullet j}.$$

Here $\mu_{\bullet j}$ is the probability that a randomly chosen data point has a "row neighbor" in column $j$ and an analogous interpretation holds for $\mu_{i\bullet}$. If column $j$ is full, then $\mu_{\bullet j} = 1$. Ordinarily we expect that most, and perhaps even all, of the $\mu_{\bullet j}$ will be small, and similarly for $\mu_{i\bullet}$.

2.1. *Random effect model.* We consider the data to have been generated by a model in which the pattern of observations has been fixed, but those observed values might have been different. That is, $Z_{ij}$ are fixed values for $i = 1, \ldots, R$ and $j = 1, \ldots, C$. The model has

(1)                        $$X_{ij} = \mu + a_i + b_j + \varepsilon_{ij},$$

where $\mu$ is an unknown fixed value and $a_i$, $b_j$ and $\varepsilon_{ij}$ are random.

In a classical random effects model (see Searle, Casella and McCulloch (1992), Chapter 5) we suppose that $a_i \sim N(0, \sigma_A^2)$, $b_j \sim N(0, \sigma_B^2)$ and $\varepsilon_{ij} \sim N(0, \sigma_E^2)$ all independently. We relax the model in several ways. By taking $a_i \sim (0, \sigma_A^2)$ for $i = 1, \ldots, R$, we mean that $a_i$ has mean 0 and variance $\sigma_A^2$ but is not necessarily normally distributed. Similarly, we suppose that $b_j \sim (0, \sigma_B^2)$ and that $\varepsilon_{ij} \sim (0, \sigma_E^2)$. We refer to this model below as the homogenous random effects model. The homogenous random effects model is a type of "superpopulation" model as often used for sampling finite populations. In a superpopulation model we suppose that a large but finite population is itself a sample from an infinite population that we want to study.

Next, there may be some measured or latent attributes making $a_i$ more variable than $a_{i'}$, for $i \neq i'$. We allow for this possibility by taking $a_i \sim (0, \sigma^2_{A(i)})$, where $\sigma^2_{A(1)}, \ldots, \sigma^2_{A(R)}$ are variances specific to the row entities. Similarly, $b_j \sim (0, \sigma^2_{B(j)})$ and $\varepsilon_{ij} \sim (0, \sigma^2_{E(i,j)})$.

The variables $a_i$, $b_j$ and $\varepsilon_{ij}$ are mutually independent. That condition can be relaxed somewhat, as described in Section 6.

The choice to model conditionally on the observed $Z_{ij}$ values is a pragmatic one. The actual mechanism generating the observations can be very complicated. It includes the possibility that any particular row or column sum might sometimes be positive and sometimes be zero. By conditioning on $Z_{ij}$, we avoid having to model unobserved entities. Also, in practice, one often finds that the smallest entities have been truncated out of the data in a preprocessing step. For example, rows might be removed if $n_{i\bullet}$ is below a cutoff like 10. A column entity that is popular with the small row entities might be seriously affected, perhaps even to the point of falling below its own cutoff level. Similarly, the large entities are sometimes removed, but for different reasons. In information retrieval, one often removes extremely common "stop words" like "and," "of" and "the" from the data. Working conditionally lets us avoid modeling this sort of truncation.

### 2.2. *Linear statistics.* We focus on a simple mean

$$\hat{\mu}_x = \frac{1}{N} \sum_i \sum_j Z_{ij} X_{ij} = \frac{1}{N} \sum_{\ell=1}^{N} X_\ell.$$

A bootstrap method that gives the correct variance for a mean can be expected to be reliable for more complicated statistics such as differences in means, other smooth functions of means and estimating equation parameters $\hat{\theta}$ defined via

$$0 = \frac{1}{N} \sum_i \sum_j Z_{ij} f(X_{ij}, \hat{\theta}) = \frac{1}{N} \sum_{\ell=1}^{N} f(X_\ell, \hat{\theta}).$$

Conversely, a bootstrap method that does not work reliably for linear statistics like a mean cannot be trusted for more complicated usages.

LEMMA 1. *Under the random effects model described above,*

$$(2) \quad V_{\mathrm{RE}}(\hat{\mu}_x) = \frac{1}{N^2} \left( \sum_{i=1}^{R} n_{i\bullet}^2 \sigma^2_{A(i)} + \sum_{j=1}^{C} n_{\bullet j}^2 \sigma^2_{B(i)} + \sum_{i=1}^{R} \sum_{j=1}^{C} Z_{ij} \sigma^2_{E(i,j)} \right).$$

*Under the homogenous random effects model,*

$$(3) \qquad V_{\mathrm{RE}}(\hat{\mu}_x) = \nu_A \frac{\sigma_A^2}{N} + \nu_B \frac{\sigma_B^2}{N} + \frac{\sigma_E^2}{N}.$$

PROOF.   Because the $a_i$, $b_i$ and $\varepsilon_{ij}$ are uncorrelated,

$$V_{\mathrm{RE}}(\hat{\mu}_x) = V_{\mathrm{RE}}\left(\frac{1}{N}\sum_{i=1}^{R}\sum_{j=1}^{C} Z_{ij}(\mu + a_i + b_j + \varepsilon_{ij})\right)$$

$$= V_{\mathrm{RE}}\left(\frac{1}{N}\sum_{i=1}^{R} n_{i\bullet} a_i\right) + V_{\mathrm{RE}}\left(\frac{1}{N}\sum_{j=1}^{C} n_{\bullet j} b_j\right) + V_{\mathrm{RE}}\left(\frac{1}{N}\sum_{i=1}^{R}\sum_{j=1}^{C} Z_{ij}\varepsilon_{ij}\right),$$

which reduces to (2). In the homogeneous case (2) further reduces to (3).   □

In the nonhomogeneous case, the unequal variance contributions for row entities are weighted proportionally to $n_{i\bullet}^2$. Thus, when the frequent entities are more variable than the others, care must be taken estimating variance components for the homogeneous model. Using a pooled estimate $\hat{\sigma}_A^2$ that weights entities equally would lead to an underestimate of the variance of $\hat{\mu}_x$.

**3. Bootstrap methods.**   We would like to bootstrap the data in such a way that the variance of the bootstrap resampled value $\hat{\mu}^*$ approximates $V_{\mathrm{RE}}(\hat{\mu}_x)$. Even better, we would like to do this without having to model the details of the random effects involved in $X_{ij}$ and without having to explicitly account for the varying $n_{i\bullet}$ values.

Here we define a naive bootstrap, which treats the data as IID, and the pigeonhole bootstrap. The latter resamples both rows and columns. Neither of these bootstraps faithfully imitates the random effects mechanism generating the data. Also, neither of them holds fixed the sample sizes $n_{i\bullet}$ and the latter does not even hold $N$ fixed. Thus both bootstraps must be tested to see whether they yield a serious systematic error.

3.1. *Naive bootstrap.*   The usual bootstrap procedure resamples the data IID from the empirical distribution. It can be reliable even when the underlying generative model fails to hold. For example, in IID regression models, resampling the cases gives reliable inferences for a regression parameter even when the regression errors have unequal variance.

It would be naive to apply IID resampling of the $N$ observed data points to the random effects setting, because the $X_{ij}$ values are not independent. Under such a naive bootstrap, $(I_\ell^*, J_\ell^*, X_\ell^*)$ is drawn independently and uniformly from the $N$ rows of $S$, for $\ell = 1, \ldots, N$. Then

$$\hat{\mu}_x^* = \frac{1}{N}\sum_{\ell=1}^{N} X_\ell^*.$$

LEMMA 2. *The expected value in the random effects model of the naive bootstrap variance of $\hat{\mu}_x^*$ is $E_{\mathrm{RE}}(V_{\mathrm{NB}}(\hat{\mu}_x^*))$, which is equal to*

$$
\begin{aligned}
(4) \quad & \frac{1}{N^2}\sum_i \sigma_{A(i)}^2 n_{i\bullet}\left(1 - \frac{n_{i\bullet}}{N}\right) \\
& + \frac{1}{N^2}\sum_j \sigma_{B(j)}^2 n_{\bullet j}\left(1 - \frac{n_{\bullet j}}{N}\right) + \frac{1}{N^2}\sum_i\sum_j Z_{ij}\sigma_{E(i,j)}^2.
\end{aligned}
$$

*Under the homogenous random effects model,*

$$
(5) \quad E_{\mathrm{RE}}(V_{\mathrm{NB}}(\hat{\mu}_x^*)) = \frac{\sigma_A^2}{N}\left(1 - \frac{\nu_A}{N}\right) + \frac{\sigma_B^2}{N}\left(1 - \frac{\nu_B}{N}\right) + \frac{\sigma_E^2}{N}.
$$

PROOF. The naive bootstrap variance of $\hat{\mu}_x^*$ is $s_x^2/N$, where $s_x^2 = (1/N)\times \sum_{\ell=1}^N (X_\ell - \hat{\mu}_x)^2$. Using the $U$-statistic formulation, we may write it as

$$
\begin{aligned}
V_{\mathrm{RE}}(\hat{\mu}_x) &= \frac{1}{2N^3}\sum_{\ell=1}^N\sum_{\ell'=1}^N (X_\ell - X_{\ell'})^2 \\
&= \frac{1}{2N^3}\sum_i\sum_j\sum_{i'}\sum_{j'} Z_{ij}Z_{i'j'}(a_i - a_{i'} + b_j - b_{j'} + e_{ij} - e_{i'j'})^2.
\end{aligned}
$$

Then under the random effects model,

$$
\begin{aligned}
& E_{\mathrm{RE}}(V_{\mathrm{NB}}(\hat{\mu}_x)) \\
& \quad = \frac{1}{2N^3}\sum_i\sum_j\sum_{i'}\sum_{j'} Z_{ij}Z_{i'j'}E(a_i - a_{i'} + b_j - b_{j'} + e_{ij} - e_{i'j'})^2 \\
& \quad = \frac{1}{2N^3}\sum_i\sum_j\sum_{i'}\sum_{j'} Z_{ij}Z_{i'j'}(1_{i\neq i'}(\sigma_{A(i)}^2 + \sigma_{A(i')}^2) \\
& \qquad\qquad\qquad\qquad\qquad + 1_{j\neq j'}(\sigma_{B(j)}^2 + \sigma_{B(j')}^2) \\
& \qquad\qquad\qquad\qquad\qquad + (1 - 1_{i=i'}1_{j=j'})(\sigma_{E(i,j)}^2 + \sigma_{E(i',j')}^2)) \\
& \quad = \frac{1}{N^2}\sum_i \sigma_{A(i)}^2 n_{i\bullet}\left(1 - \frac{n_{i\bullet}}{N}\right) + \frac{1}{N^2}\sum_j \sigma_{B(j)}^2 n_{\bullet j}\left(1 - \frac{n_{\bullet j}}{N}\right) \\
& \qquad + \frac{1}{N^2}\sum_i\sum_j Z_{ij}\sigma_{E(i,j)}^2. \qquad\qquad\qquad\qquad\qquad\qquad\square
\end{aligned}
$$

The homogenous case shows the differences clearly. The error term $\sigma_E^2$ gets accounted for correctly, but not the other terms. Where the row variable really contributes $\nu_A\sigma_A^2/N$ to the variance, the naive bootstrap only captures $\sigma_A^2(1 - \nu_A/N)/N$ of it. It underestimates this variance by a factor of $\nu_A/(1-$ $\nu_A/N) \approx \nu_A$, which may be substantial. The variance due to the column variables is similarly under-estimated in the naive bootstrap.

3.2. *Pigeonhole bootstrap.* The naive bootstrap fails because it ignores similarities between elements of the same row and/or column. A more principled bootstrap would be based on estimating the component parts of the random effects model, and resampling from it. However, finding a good way to estimate such a model can be burdensome. Normal theory random effects models for severely unbalanced data sets are hard to fit. More worryingly, the data may depart seriously from normality, the variances $\sigma^2_{A(i)}$ might be nonconstant, and could even be correlated somehow with the sample sizes $n_{i\bullet}$.

The pigeonhole bootstrap is named for a model used by Cornfield and Tukey (1956) to study balanced anovas with fixed, mixed and random effects. We place the data into an $R \times C$ matrix, resample a set of rows, resample a set of columns, and take the intersections as the bootstrapped data set. The original pigeonhole model involved sampling without replacement. The pigeonhole bootstrap samples with replacement. The appeal of the pigeonhole bootstrap is that it generates a resampled data set with values taken from the original data. There is no need to form synthetic combinations of row and column entities that were never observed together, nor response values $X_{ij}$ that were not observed.

Formally, in the pigeonhole bootstrap we sample rows $r_i^*$ IID from $U\{1, \dots, R\}$ for $i = 1, \dots, R$ and columns $c_j^*$ IID from $U\{1, \dots, C\}$ for $j = 1, \dots, C$. Rows and columns are sampled independently.

The resampled data set has $Z_{ij}^* = Z_{r_i^* c_j^*}$ and when $Z_{ij}^* = 1$, we take $X_{ij}^* = X_{r_i^* c_j^*}$. The bootstrap sample sizes are $n_{i\bullet}^* = \sum_{j=1}^{C} Z_{ij}^*$, $n_{\bullet j}^* = \sum_{i=1}^{R} Z_{ij}^*$ and $N^* = n_{\bullet\bullet}^* = \sum_{i=1}^{R} \sum_{j=1}^{C} Z_{ij}^*$.

The bootstrap process above is repeated independently, some number $B$ of times.

Row $i = 1$ in $Z_{ij}^*$ does not ordinarily correspond to the same entity as row $i = 1$ in $Z_{ij}$. Should we want to keep track of the number of times that original row entity $r$ appeared in the resampled data, we would use

$$\widetilde{n}_{r\bullet}^* = \sum_{i=1}^{R} \sum_{j=1}^{C} Z_{r_i^* c_j^*} \times 1_{r_i^* = r} = \sum_{i=1}^{R} n_{i\bullet}^* 1_{r_i^* = r}.$$

Similarly, column $c$ appears $\widetilde{n}_{\bullet c}^* = \sum_{j=1}^{C} n_{\bullet j}^* 1_{c_j^* = c}$ times among the resampled values.

**4. Variances.** Here we develop the variance formulas for pigeonhole bootstrap sampling. It is convenient to consider the entities as belonging to a large but finite set. Then we may work with population totals, a concept that makes no sense for an infinite pool of entities.

There are two uncertainties in a bootstrap variance estimate. One is sampling uncertainty, the variance under bootstrapping $B$ times of our variance estimate. The other is systematic uncertainty, the difference between the expectation of our bootstrap variance estimate and the variance we wish to estimate. We focus on the latter because the resampling model differs from the one we assume has generated the data. The former issue can be helped by increasing $B$, and will be less severe for large $N$. It will be possible to construct examples where sampling fluctuations dominate, but we do not consider that case here. McCullagh (2000) also chose to compare expected bootstrap variance to sampling variance.

4.1. *Variance of totals in pigeonhole bootstrap.* The total value of $X$ over the sample is $T_x = \sum_i \sum_j Z_{ij} X_{ij}$. In bootstrap sampling, the total of $X^*$ is $T_x^* = \sum_i \sum_j Z_{ij}^* X_{ij}^*$.

LEMMA 3. *Under pigeonhole bootstrap sampling,*

$$E_{\mathrm{PB}}(T_x^*) = T_x \quad and \quad E_{\mathrm{PB}}(N^*) = N.$$

PROOF. First, $E(T_x^*) = \sum_{i=1}^R \sum_{j=1}^C E_B(Z_{ij}^* X_{ij}^*) = RC E_{\mathrm{PB}}(Z_{11}^* X_{11}^*)$ by symmetry. The first result then follows because $E_{\mathrm{PB}}(Z_{11}^* X_{11}^*) = 1/(RC) \sum_i \sum_j Z_{ij} \times X_{ij}$. The second result follows on putting $N = T_z$ and $N^* = T_z^*$ and considering the case $X_{ij} = Z_{ij}$. □

THEOREM 1. *Under pigeonhole bootstrap sampling,*

$$V_{\mathrm{PB}}(T_x^*) = \left( \frac{1}{RC} - \frac{1}{R} - \frac{1}{C} \right) T_x^2 + \left( 1 - \frac{1}{C} \right) \sum_i T_{xi\bullet}^2$$

$$+ \left( 1 - \frac{1}{R} \right) \sum_j T_{x\bullet j}^2 + \sum_i \sum_j Z_{ij} X_{ij}^2,$$

*where* $T_{xi\bullet} = \sum_{j=1}^C Z_{ij} X_{ij}$ *and* $T_{x\bullet j} = \sum_{i=1}^R Z_{ij} X_{ij}$.

PROOF. We write

$$E_{\mathrm{PB}}((T_x^*)^2) = \sum_i \sum_j \sum_{i'} \sum_{j'} E_{\mathrm{PB}}(Z_{ij}^* X_{ij}^* Z_{i'j'}^* X_{i'j'}^*)$$

and then split the sum into four cases depending on whether $i = i'$ or not and whether $j = j'$ or not. By symmetry, we only need to consider $i$, $j$, $i'$ and $j'$ equal to 1 or 2. Thus,

$$
\begin{aligned}
E_{\mathrm{PB}}((T_x^*)^2) &= R(R-1)C(C-1)E_{\mathrm{PB}}(Z_{11}^*X_{11}^*Z_{22}^*X_{22}^*) \\
&\quad + RC(C-1)E_{\mathrm{PB}}(Z_{11}^*X_{11}^*Z_{12}^*X_{12}^*) \\
&\quad + R(R-1)CE_{\mathrm{PB}}(Z_{11}^*X_{11}^*Z_{21}^*X_{21}^*) \\
&\quad + RCE_{\mathrm{PB}}(Z_{11}^*X_{11}^*Z_{11}^*X_{11}^*) \\
&= R(R-1)C(C-1)\frac{T_x^2}{R^2C^2} + RC(C-1)\frac{1}{RC^2}\sum_i T_{xi\bullet}^2 \\
&\quad + R(R-1)C\frac{1}{R^2C}\sum_j T_{x\bullet j}^2 + RC\frac{1}{RC}\sum_i\sum_j Z_{ij}X_{ij}^2,
\end{aligned}
$$

which, combined with Lemma 3, yields the result.  □

4.2. *Ratio estimation.* The simple mean of $X$ is $\hat{\mu}_x = T_x/N$. Under bootstrap resampling, we generate $\hat{\mu}_x^* = T_x^*/N^*$. The mean and variance of $\hat{\mu}_x^*$ are complicated because they are ratios. The bootstrap sample size $N^*$, appearing in the denominator, is not constant.

There is a standard way to handle ratio estimators in sampling theory [Cochran (1977)]. It amounts to use of the delta method. The approximate variance of $\hat{\mu}_x$ using ratio estimation takes the form

$$
(6) \qquad \hat{V}_{\mathrm{PB}}(\hat{\mu}_x^*) = \frac{1}{N^2}E_{\mathrm{PB}}\left(\left(T_x^* - \frac{T_x}{N}N^*\right)^2\right).
$$

THEOREM 2. *Under pigeonhole bootstrap sampling,*

$$
\hat{V}_{\mathrm{PB}}(\hat{\mu}_x^*) = \frac{1}{N^2}\left[\left(1-\frac{1}{C}\right)\sum_i n_{i\bullet}^2(\bar{X}_{i\bullet}-\hat{\mu}_x)^2 + \left(1-\frac{1}{R}\right)\sum_j n_{\bullet j}^2(\bar{X}_{\bullet j}-\hat{\mu}_x)^2 \right. \\
\left. + \sum_i\sum_j Z_{ij}(X_{ij}-\hat{\mu}_x)^2\right],
$$

*where $\bar{X}_{i\bullet} = \sum_{j=1}^C Z_{ij}X_{ij}/n_{i\bullet}$ and $\bar{X}_{\bullet j} = \sum_{i=1}^R Z_{ij}X_{ij}/n_{\bullet j}$.*

PROOF. Let $Y_{ij} = X_{ij} - T_x Z_{ij}/N$ when $Z_{ij}=1$. Then $T_y = T_x - T_x T_z/N = 0$. Also, $T_y^* = T_x^* - T_x N^*/N$ and so $E_{\mathrm{PB}}((T_x^* - T_x N^*/N)^2) = V_{\mathrm{PB}}(T_y^*)$.

From Theorem 1 applied to $Y$ instead of $X$ we have

$$
\hat{V}_{\mathrm{PB}}(\hat{\mu}_x^*) = \frac{1}{N^2}\left[\left(1-\frac{1}{C}\right)\sum_i T_{yi\bullet}^2 + \left(1-\frac{1}{R}\right)\sum_j T_{y\bullet j}^2 + \sum_i\sum_j Z_{ij}Y_{ij}^2\right].
$$

We find that $X_{ij} - Z_{ij}T_x/N = X_{ij} - \hat{\mu}_x$ whenever $Z_{ij} = 1$, and so

$$\sum_i T_{yi\bullet}^2 = \sum_i \left( \sum_j Z_{ij}(X_{ij} - Z_{ij}T_x/N) \right)^2 = \sum_i \left( \sum_j Z_{ij}(X_{ij} - \hat{\mu}_x) \right)^2$$

$$= \sum_i n_{i\bullet}^2 (\bar{X}_{i\bullet} - \hat{\mu}_x)^2$$

and a similar expression holds for $\sum_i T_{yi\bullet}^2$. Also,

$$\sum_i \sum_j Z_{ij} Y_{ij}^2 = \sum_i \sum_j Z_{ij}(X_{ij} - \hat{\mu}_x)^2. \qquad \square$$

Next we study the expected value under the random effects model of the bootstrap variance $\hat{V}_{\mathrm{PB}}(\hat{\mu}_x^*)$. We get an exact but lengthy formula and then apply simplifications.

THEOREM 3. *The expected value under the random effects model of the pigeonhole variance is*

$$E_{\mathrm{RE}}(\hat{V}_{\mathrm{PB}}(\hat{\mu}_x^*)) = \frac{1}{N^2} \left[ \sum_i \sigma_{A(i)}^2 \lambda_i^A + \sum_j \sigma_{B(j)}^2 \lambda_j^B + \sum_i \sum_j Z_{ij} \sigma_{E(i,j)}^2 \lambda_{i,j}^E \right],$$

*where*

$$\lambda_i^A = \left( 1 - \frac{1}{C} \right) n_{i\bullet}^2 \left( 1 - 2\frac{n_{i\bullet}}{N} + \frac{\nu_A}{N} \right) + \left( 1 - \frac{1}{R} \right) \left( n_{i\bullet} - 2\mu_{i\bullet} n_{i\bullet} + \frac{\nu_B n_{i\bullet}^2}{N} \right)$$
$$+ n_{i\bullet} \left( 1 - \frac{n_{i\bullet}}{N} \right)^2 + \frac{n_{i\bullet}^2}{N^2}(N - n_{i\bullet}),$$

$$\lambda_j^B = \left( 1 - \frac{1}{R} \right) n_{\bullet j}^2 \left( 1 - 2\frac{n_{\bullet j}}{N} + \frac{\nu_B}{N} \right) + \left( 1 - \frac{1}{C} \right) \left( n_{\bullet j} - 2\mu_{\bullet j} n_{\bullet j} + \frac{\nu_A n_{\bullet j}^2}{N} \right)$$
$$+ n_{\bullet j} \left( 1 - \frac{n_{\bullet j}}{N} \right)^2 + \frac{n_{\bullet j}^2}{N^2}(N - n_{\bullet j}),$$

*and for $Z_{ij} = 1$,*

$$\lambda_{i,j}^E = \left( 1 - \frac{1}{C} \right) \left( 1 - 2\frac{n_{i\bullet}}{N} + \frac{\nu_A}{N} \right) + \left( 1 - \frac{1}{R} \right) \left( 1 - 2\frac{n_{\bullet j}}{N} + \frac{\nu_B}{N} \right) + 1 - \frac{1}{N}.$$

PROOF. See the [Appendix](#).

The variance expression above is unwieldy. We use the notation $\approx$ to indicate that terms like $1/R$, $1/C$, $n_{i\bullet}/N$, $n_{\bullet j}/N$, $\nu_A/N$, $\nu_B/N$, $\mu_{i\bullet}$ and $\mu_{\bullet j}$ are considered negligible compared to 1, as they usually are in the motivating applications. We do not suppose that $n_{i\bullet}/n_{i\bullet}^2$ is negligible because some or even most of the $n_{i\bullet}$ could be small. Under these conditions,

$$\lambda_i^A \approx n_{i\bullet}^2 + 2n_{i\bullet},$$

$$\lambda_j^B \approx n_{\bullet j}^2 + 2n_{\bullet j} \quad \text{and for } Z_{ij} = 1,$$

$$\lambda_{i,j}^E \approx 3.$$

COROLLARY 1.

$$E_{\text{RE}}(\hat{V}_{\text{PB}}(\hat{\mu}_x^*))$$

(7)
$$\approx \frac{1}{N^2}\Bigg[\sum_i \sigma_{A(i)}^2(n_{i\bullet}^2 + 2n_{i\bullet})$$

$$+ \sum_j \sigma_{B(j)}^2(n_{\bullet j}^2 + 2n_{\bullet j}) + 3\sum_i \sum_j Z_{ij}\sigma_{E(i,j)}^2\Bigg],$$

*and under the homogenous random effects model,*

$$E_{\text{RE}}(\hat{V}_{\text{PB}}(\hat{\mu}_x^*)) \approx \frac{1}{N}(\sigma_A^2(\nu_A + 2) + \sigma_B^2(\nu_B + 2) + 3\sigma_{E(i,j)}^2).$$

PROOF.

$$E_{\text{RE}}(\hat{V}_{\text{PB}}(\hat{\mu}_x^*))$$

$$= \frac{1}{N^2}\Bigg[\sum_i \sigma_{A(i)}^2\lambda_i^A + \sum_j \sigma_{B(j)}^2\lambda_j^B + \sum_i \sum_j Z_{ij}\sigma_{E(i,j)}^2\lambda_{i,j}^E\Bigg]$$

$$\approx \frac{1}{N^2}\Bigg[\sum_i \sigma_{A(i)}^2(n_{i\bullet}^2 + 2n_{i\bullet})$$

$$+ \sum_j \sigma_{B(j)}^2(n_{\bullet j}^2 + 2n_{\bullet j}) + 3\sum_i \sum_j Z_{ij}\sigma_{E(i,j)}^2\Bigg].$$

The specialization to the homogenous case follows easily.  □

The variance contribution from $\varepsilon_{ij}$ is overestimated by a factor of 3 in the pigeonhole bootstrap. In the homogenous case this overestimate is not important when $\sigma_E^2 \ll \max(\nu_A\sigma_A^2, \nu_B\sigma_B^2)$. When $\nu_A$ and $\nu_B$ are large, it only takes a small amount of variation in $\sigma_A^2$ and $\sigma_B^2$ to make $\sigma_E^2$ unimportant. A similar conclusion follows for the inhomogenous case in terms of appropriately weighted averages of the variances.

The average value of the variance in equation (7) tracks very closely with the desired random effects variance of $\hat{\mu}_x$ given by (2), even when the effects are heteroscedastic. Where the latter has $n_{i\bullet}^2$, the former has $n_{i\bullet}^2 + 2n_{i\bullet}$, and similarly for $n_{\bullet j}^2$. Outside of extreme cases $\sum_i n_{i\bullet}\sigma_{A(i)}^2 \ll \sum_i n_{i\bullet}^2\sigma_{A(i)}^2$.

In some applications some or many of the $\mu_{i\bullet}$ and $\mu_{\bullet j}$ may be nontrivially large. In recommender settings, a small number of books or movies may have been rated by a large fraction of people, or some people may have rated an astonishingly large number of items. In information retrieval, some terms might appear in most documents, as, for example, when we choose to retain the stop words. Under such conditions, we get a slight variance reduction:

$$\lambda_i^A \approx n_{i\bullet}^2 + 2(1-\mu_{i\bullet})n_{i\bullet}$$

and

$$\lambda_j^B \approx n_{\bullet j}^2 + 2(1-\mu_{\bullet j})n_{\bullet j},$$

while $\lambda_{i,j}^E$ remains approximately 3. But when $\mu_{i\bullet}$ is not small, then we may reasonably expect $n_{i\bullet}$ to be large and $n_{i\bullet}^2 \gg n_{i\bullet}$. Thus, the approximation in Corollary 1 is still appropriate.

4.3. *Mean consistency.* The expression $\approx$ conveys what we ordinarily expect to be the important terms. We find $E_{\mathrm{RE}}\hat{V}_{\mathrm{PB}}(\hat{\mu}_x^*) \approx V_{\mathrm{RE}}(\hat{\mu}_x)$ or $E_{\mathrm{RE}}\hat{V}_{\mathrm{PB}}(\hat{\mu}_x^*)/V_{\mathrm{RE}}(\hat{\mu}_x) \approx 1$. However, the earlier section left open the possibility of extreme cases where $\sum_i n_{i\bullet}\sigma_{A(i)}^2$ was not negligible compared to $\sum_i n_{i\bullet}^2\sigma_{A(i)}^2$. For example, suppose that $\sigma_{A(i)}^2 = 0$ for all $i$ with $n_{i\bullet} > 1$. Then $(n_{i\bullet}^2 + 2n_{i\bullet})\sigma_{A(i)}^2 = 3n_{i\bullet}^2\sigma_{A(i)}^2$ and the pigeonhole bootstrap could essentially triple the variance contribution from the row entities.

To formulate "mean consistency" of $\hat{V}_{\mathrm{PB}}(\hat{\mu}_x^*)$ more carefully, we define

$$(8) \qquad \epsilon_N = \max\left(\frac{1}{R}, \frac{1}{C}, \frac{\nu_A}{N}, \frac{\nu_B}{N}, \frac{1}{\nu_A}, \frac{1}{\nu_B}, \max_i \frac{n_{i\bullet}}{N}, \max_j \frac{n_{\bullet j}}{N}\right),$$

and work in the limit as $N \to \infty$ with $\epsilon_N \to 0$.

Arranging the terms in each $\lambda$, we get

$$\lambda_i^A = n_{i\bullet}^2(1 + O(\epsilon_N)) + 2n_{i\bullet}(1-\mu_{i\bullet})(1 + O(\epsilon_N)),$$
$$\lambda_j^B = n_{\bullet j}^2(1 + O(\epsilon_N)) + 2n_{\bullet j}(1-\mu_{\bullet j})(1 + O(\epsilon_N))$$

and

$$\lambda_{i,j}^E = 3 + O(\epsilon_N),$$

where the implied constant in all the $O(\epsilon_N)$ terms is independent of $i$ and $j$.

To rule out pathological heteroscedasticity, we suppose that

$$0 < m_A \leq \sigma_{A(i)}^2 \leq M_A < \infty,$$
$$0 < m_B \leq \sigma_{B(j)}^2 \leq M_B < \infty$$

and

$$0 < m_E \le \sigma^2_{E(i,j)} \le M_E < \infty$$

holds for all $1 \le i \le R = R(N)$ and $1 \le j \le C = C(N)$.

THEOREM 4. *Suppose that $\sigma^2_{A(i)}$, $\sigma^2_{B(j)}$ and $\sigma^2_{E(i,j)}$ obey the bounds above, and that $\epsilon_N \to 0$ as $N \to \infty$. Then*

$$\frac{E(V_{\mathrm{PB}}(\hat\mu_x)) - V_{\mathrm{RE}}(\hat\mu_x)}{V_{\mathrm{RE}}(\hat\mu_x)} = O(\epsilon_N).$$

PROOF. Gathering up the pieces,

$$\frac{E(V_{\mathrm{PB}}(\hat\mu_x)) - V_{\mathrm{RE}}(\hat\mu_x)}{V_{\mathrm{RE}}(\hat\mu_x)} = O(\epsilon_N) + 2(1 + O(\epsilon_N)) \times \rho_N,$$

where

$$\rho_N = \frac{\sum_i n_{i\bullet}(1 - \mu_{i\bullet})\sigma^2_{A(i)} + \sum_j n_{\bullet j}(1 - \mu_{\bullet j})\sigma^2_{B(j)} + \sum_i \sum_j Z_{ij}\sigma^2_{E(i,j)}}{\sum_i n^2_{i\bullet}\sigma^2_{A(i)} + \sum_j n^2_{\bullet j}\sigma^2_{B(j)} + \sum_i \sum_j Z_{ij}\sigma^2_{E(i,j)}}.$$

The numerator of $\rho_N$ lies between $N m_E$ and $N(M_A + M_B + M_E)$, while the denominator is at least $N(\nu_A m_A + \nu_B m_B + m_E)$. Therefore,

$$0 \le \rho_N \le \frac{M_A + M_B + M_E}{\nu_A m_A + \nu_B m_B + m_E} = O(\epsilon_N). \qquad \square$$

4.4. *Covariances.* The variance expressions in this paper generalize in an unsurprising way to covariances of pairs of responses. The simplest way to express this is to suppose that $X_{ij} \in \mathbb{R}^d$. Then we may generalize $\sigma^2_{A(i)}$, $\sigma^2_{B(j)}$ and $\sigma^2_{E(i,j)}$ to be $d \times d$ covariance matrices. The variance formulas go through as above. Expressions like $\sum_i n_{i\bullet}\sigma^2_{A(i)} \ll \sum_i n^2_{i\bullet}\sigma^2_{A(i)}$ then mean that $\sum_i n_{i\bullet}u'\sigma^2_{A(i)}u \ll \sum_i n^2_{i\bullet}u'\sigma^2_{A(i)}u$ for all $u \in \mathbb{R}^d$ with $\|u\| = 1$.

**5. Netflix movie ratings example.** As an example of a small effect near the uncertainty level, we consider the day of the week effect in movie ratings for the Netflix data. This data set is described and is available from **http://www.netflixprize.com/index**. It has 100,480,507 ratings on 17,770 movies from 480,189 customers. As mentioned above, $\nu$ for customers is about 646 and $\nu$ for movies is about 56,200. The number of ratings per customer ranges from 1 to 17,653. The number of ratings for movies ranges from 3 to 232,944.

5.1. *Day of the week effect.* It would be interesting to examine the effects of demographic variables on movie ratings, but for privacy purposes those are not included in the data. The data set does, however, supply a date for each rating. For each day of the week, we may find the average movie rating given out. The smallest value is 3.595808 for Tuesdays and the largest is 3.616449 for Sundays. The day of the week effect is very small.

Perhaps the movie ratings given out on Sunday do tend to be larger than those given out on Sunday. If so, we might investigate whether this arises from a different mix of movies being rated that day, a different set of customers rating on that day, or some subtle interaction. But before doing such followup, we should check whether the difference might just be a sampling fluctuation. For such a large data set, sampling fluctuations are expected to be small. But the observed effect is also quite small, and the sampling fluctuations include random effects from movies and customers that can make them much larger than we were used to in the IID setting.

Figure 1 shows results for 10 pigeonhole bootstrap samples. In each sample the means for all 7 days of the week were recorded. There is clearly a bias in the bootstrap resampling. The average score on any given day in resampling is a ratio estimate of the total of all scores given for that day divided by the number of ratings for that day. The bias is very small in absolute terms, but not compared to the pigeonhole bootstrap standard deviation.

A day versus day comparison is more interesting than the absolute level for a given day. To compare Tuesday and Sunday, we look at the 10 paired average scores. These are shown in Figure 2. The solid point is 0.0206 units above the forty-five degree line, indicating that Sunday scores average that much higher than Tuesday scores. The resampled points average 0.0214 units above the line. This is very close to the sample difference. The biases for the two days' scores have almost completely cancelled out, so that some resampled points are farther from the line than the original point while others are closer. The average resampled difference in means is about 8.16 times as large as the standard deviation of the resampled differences. The 10 bootstrap differences are independent, and should be nearly normal because of the large sample sizes involved. Then $\Pr(|t_{(9)}| \geq 8.16) \doteq 1.88 \times 10^{-5}$. This is small enough for us to conclude that the difference is real, even if we take account of having selected the most significant of all 21 day to day comparisons.

5.2. *Parameters and hypothesis.* This section makes clear what hypothesis is being tested by the bootstrap analysis and what are the underlying parameters. Then it looks at how well the random effects model might fit the present setting.

Let $Y_{ij}$ be the score for a movie and let $Z_{ij}$ be an indicator that it was observed. Introduce binary covariates $D_{ij}^{\text{Tue}}$ taking the value 1 if and only if the $ij$ measurement happened on Tuesday. Similarly, let $D_{ij}^{\mathrm{Sun}}$ be the day of week indicator for Sunday.

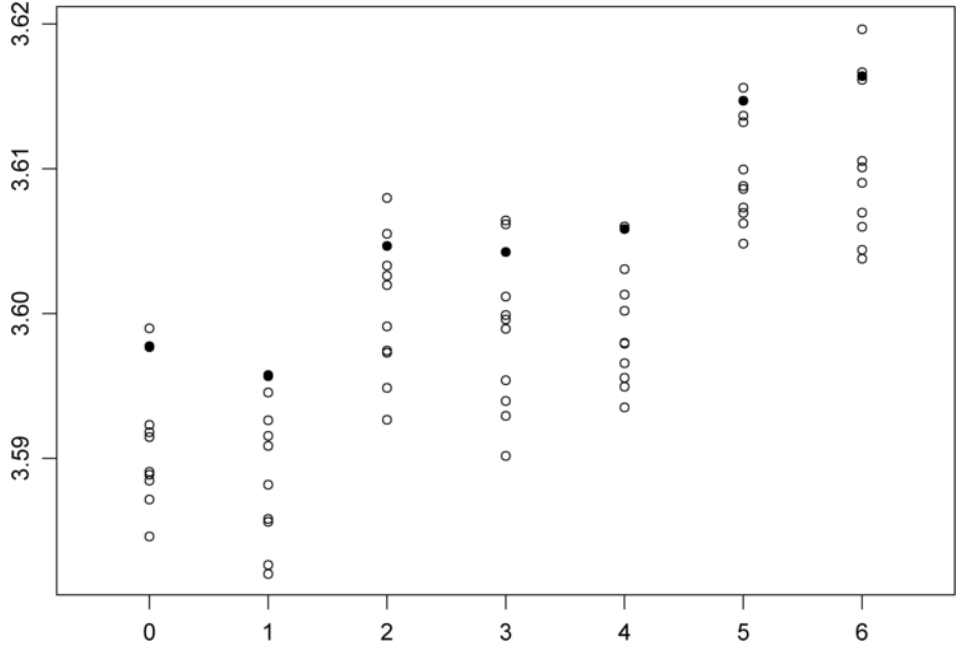

FIG. 1. *The horizontal axis depicts the day of the week from Monday at 0 to Sunday at 6. The vertical axis has average movie rating scores. For each day the solid dot shows that day's average movie rating in the original data set. The open dots show the average in each of 10 pigeonhole bootstrap samples.*

The sample average for Tuesday is

$$(9) \qquad \hat{\mu}_{\mathrm{Tue}} = \frac{\sum_{ij} Z_{ij} D_{ij}^{\mathrm{Tue}} Y_{ij}}{\sum_{ij} Z_{ij} D_{ij}^{\mathrm{Tue}}}$$

and $\hat{\mu}_{\mathrm{Sun}}$ is defined similarly. We interpret $\hat{\mu}_{\mathrm{Tue}}$ above as an estimate of

$$(10) \qquad \mu_{\mathrm{Tue}} = \frac{E_{\mathrm{RE}}(\sum_{ij} Z_{ij} D_{ij}^{\mathrm{Tue}} Y_{ij})}{E_{\mathrm{RE}}(\sum_{ij} Z_{ij} D_{ij}^{\mathrm{Tue}})},$$

so that $\mu_{\mathrm{Tue}}$ is the solution to $E_{\mathrm{RE}}(\sum_{ij} Z_{ij} D_{ij}^{\mathrm{Tue}}(Y_{ij} - \mu_{\mathrm{Tue}})) = 0$.

The null hypothesis $H_0$ being tested by the pigeonhole bootstrap analysis is that $\mu_{\mathrm{Sun}} - \mu_{\mathrm{Tue}} = 0$.

In bootstrapping $\hat{\mu}_{\mathrm{Tue}}$ plays the role of the parameter and $\hat{\mu}_{\mathrm{Tue}}^*$ plays that of the estimate. The parameter $\mu_{\mathrm{Tue}}$ is well defined so long as $\Pr(\sum_{ij} Z_{ij} D_{ij}^{\mathrm{Tue}} = 0) > 0$. We have neglected the possibility that the denominator in $\hat{\mu}_{\mathrm{Tue}}^*$ is zero. In practice, one might add a small constant to the denominator of each $\hat{\mu}_{\mathrm{Tue}}^*$, or condition on the denominator being positive. Such small resampled denominators are, however, a sign that the delta method approximation for the variance of the ratio may be inaccurate.

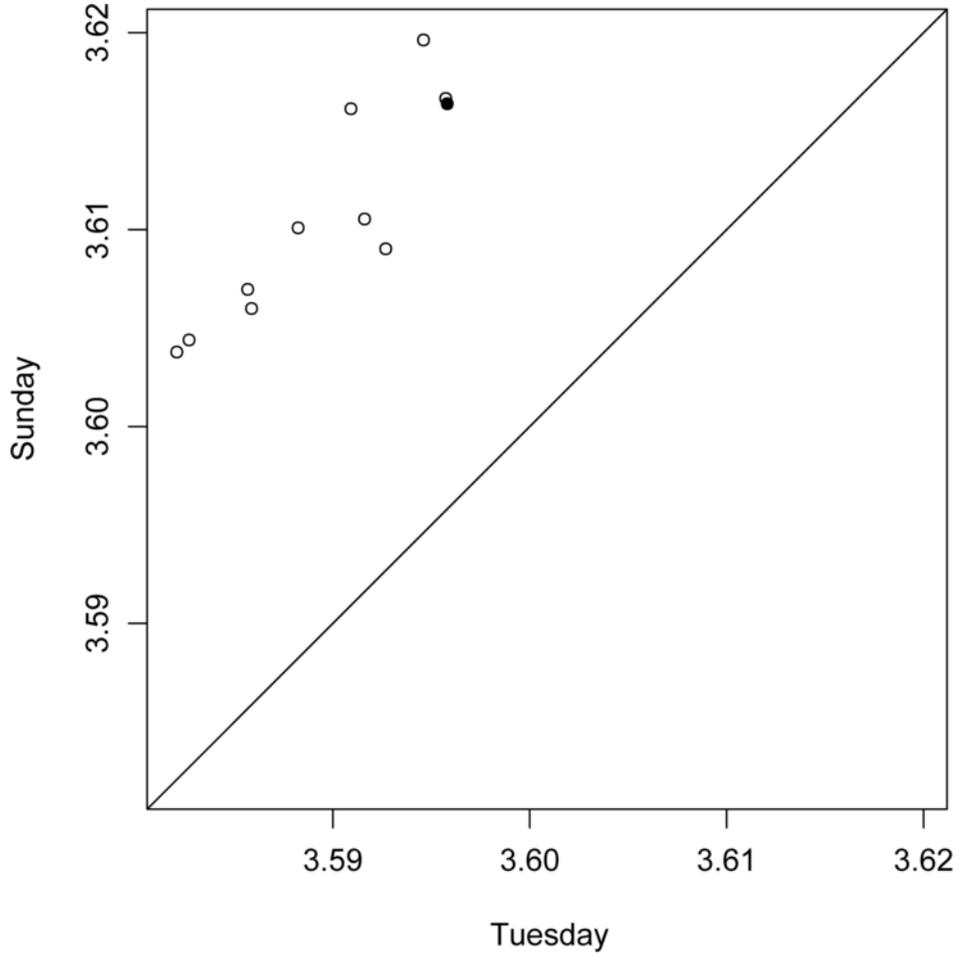

Fig. 2. *The horizontal axis shows the average movie rating given on Tuesday. The vertical axis shows Sunday. The open circles are from 10 pigeonhole bootstrap samples. The solid point is from the original data. For each day the solid dot shows that day's average movie rating in the original data set. The open dots show the average in each of 10 bootstrap samples.*

The simplest path between the random effects model and the data set is to suppose that the triple $(Y_{ij}, D_{ij}^{\mathrm{Tue}}, D_{ij}^{\mathrm{Sun}})$ approximately follows a crossed random effects model of the form studied in this paper. Such a model is somewhat unnatural though, because two of the variables are binary.

All we need, however, is that the first term in the linearization of the test statistic follows approximately a crossed random effects model. The linearized statistic is not binary.

Dropping the subscript and superscript for Tuesday, and writing the ratio in (9) as $T_{zdy}/T_{zd}$, the delta method linearization of this ratio is

$$\mu + \frac{T_{zdy} - E(T_{zdy})}{E(T_{zd})} - \frac{T_{zd} - E(T_{zd})}{E(T_{zd})^2}$$
$$= \text{const} + \sum_{ij} Z_{ij} D_{ij}(Y_{ij} E(T_{zd})^{-1} - E(T_{zd})^{-2}).$$

The linearization for Tuesdays takes six different values, corresponding to five different values when $D_{ij}^{\text{Tue}} = 1$ and one value for all cases with $D_{ij}^{\text{Tue}} = 0$. After linearizing the Sunday versus Tuesday difference, there are eleven different values arising as five for Sunday ratings, five for Tuesday ratings, and one for ratings made on the other five days.

5.3. *A deeper look into the data.* Sunday ratings are about 0.02 points higher than Tuesday ratings. This difference is small, but statistically significant, even allowing for random customer and movie effects.

What follows is an informal data analytic exploration of the nature of this discrepancy. Making the analysis formal would take us somewhat beyond the results proved here.

Several explanations for the day of week effect are plausible. One is that the harder to please customers rate more often on Tuesday. A second is that a given customer making a rating supplies a lower rating if the day is Tuesday. There are also versions of these explanations centered on movies. The movies rated on Tuesday might tend to be less popular, or a given movie getting rated on a Tuesday might tend to get a lower rating than on a Sunday.

A simple proxy for how well a movie is liked is the average score it gets. Similarly, the generosity level of a customer may be judged by the average score that he or she gives. There were 17,784,852 Tuesday ratings and the average over these ratings of our simple movie score is 3.600 (rounded to the nearest 1/1000th), slightly higher than the simple average of all Tuesday scores. The comparable numbers for Sunday are 10,730,350 and 3.616. Therefore, it appears that slightly more popular movies are being rated on Sundays than on Tuesdays. This simple analysis yields an estimated gap of 0.016 compared to the observed gap of 0.020. A customer-based version of this analysis yields a gap of only 0.010 (from 3.610 versus 3.600).

By this analysis, the effect seems to be due slightly more to which movies are being ranked than to who is doing the ranking. If we add both effects, we get a predicted gap of 0.026. This value is higher than what was observed, indicating that some amount of double counting is taking place. This double counting is consistent with a strong feature in the data. Popular movies get more ratings and also higher ratings. Very active customers give more ratings, have a harder time restricting themselves to just the popular movies, and, not surprisingly, tend to give out lower ratings. Thus, knowing that Tuesday has the less popular movies already leads one to suspect it will have the busier and, hence, less generous customers.

Another analysis looks at customers who gave ratings on both Tuesday and Sunday. For each such customer we can measure their average Sunday rating and subtract their average Tuesday rating. This gives each customer a Sunday versus Tuesday differential. The mean over customers of this differential is 0.016. Perhaps coincidentally, this matches the average movie effect. The mean over movies of a comparable movie differential is $-0.008$. It has an unexpected sign, meaning that by this measure Sunday scores are lower than Tuesday scores. A more proper analysis uses a weighted average of customers or movies, with weights depending on how many data points they contribute. The proper weighting may be a matter of debate, but, for simplicity, we use the harmonic mean of $n_{\mathrm{Tue}}$ and $n_{\mathrm{Sun}}$ where these are the number of Tuesday and Sunday ratings made by the customer. Now the weighted mean differential is 0.009. Taking this at face value, customers seem to be slightly harder to please on Tuesday. A similar weighted analysis by movies gives a differential of 0.011, so movies tend to get lower scores on Tuesday.

The pattern in the differentials is somewhat more subtle than the analysis above describes. For a given customer, let $Y$ denote the average of their Sunday scores minus the average of their Tuesday scores and let $X$ denote the simple average of those two average scores. A plot of $Y$ versus $X$ has a great many data points, but a spline smooth, using the harmonic mean weights described above shows a pattern. Generous customers are even more generous on Sundays than Tuesdays. But hard to please customers give even lower ratings on Sundays than Tuesdays.

In other words, the customers are slightly more extreme on Sundays than they are on Tuesdays. Because high scores are more common, this raises the average score on Sunday versus Tuesday. A comparable analysis by movie shows that unpopular movies get even lower scores on Tuesdays, popular movies get about the same score on both days, and intermediate movies get somewhat higher scores on Tuesdays. These curves are shown in Figure 3.

The informal data analysis above gives some support to all four explanations offered. Tuesday appears to get more of the tougher customers and the weaker movies. Furthermore, a given customer or movie seems to result in a lower score if the rating is made on Tuesday. From the figure we see that the within customer or within movie day of the week effect can be positive or negative and may be much larger than 0.002 in absolute value.

**6. Discussion.** Ultimately we would like to have a trustworthy bootstrap analysis for elaborate methods such as the spectral biclustering procedure of Dhillon (2001), among others. Running five or ten repeats would give insight as to which features of the analysis remain stable and which ones might be idiosyncratic to the data set at hand. There is a large complexity gap between the output of such methods and the simple mean considered here. It does, however, seem reasonable to rule out methods that cannot handle the global mean and focus further research on one that does.

We would also wish to have a bootstrap that works under more flexible settings than the additive random effects model in (1). The rest of this discussion presents two simple generalizations of (1) where the pigeonhole bootstrap can be applied and then discusses the issue of bias, and the difference between fixed and random $Z_{ij}$ models.

6.1. *Relaxing independence.*   It is not hard to see that the same variance results arise if the $a_i$, $b_j$ and $\varepsilon_{ij}$ are simply uncorrelated and not necessarily independent. This helps us in settings where $X_{ij}$ must be in a restricted set.

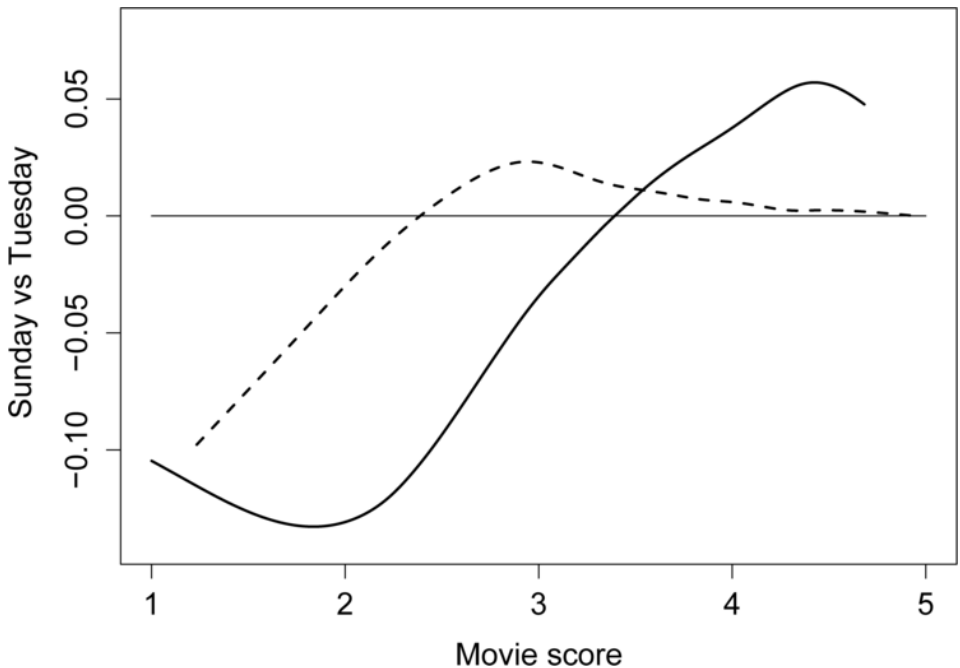

FIG. 3.   *These curves show how the Sunday versus Tuesday rating difference varies with the popularity of movies. The solid curve shows an analysis by customers, the dashed curve shows an analysis by movies. Suppose that a customer makes $n_{\mathrm{Tue}} \geq 1$ ratings on Tuesday with average value $\bar{Y}_{\mathrm{Tue}}$ and similarly for Sunday. Then the solid curve is a spline smooth on 8 degrees of freedom of $\bar{Y}_{\mathrm{Sun}} - \bar{Y}_{\mathrm{Tue}}$ versus $(\bar{Y}_{\mathrm{Sun}} + \bar{Y}_{\mathrm{Tue}})/2$ over customers, with weights $2/(1/n_{\mathrm{Sun}} + 1/n_{\mathrm{Sun}})$. The dashed curve is computed the same way, using counts and averages per movie.*

For example, if the $X_{ij}$ values can only belong to a finite set, such as movie ratings $\{1, 2, 3, 4, 5\}$, then given $a_i$ and $b_j$ there are only 5 allowable values for $\varepsilon_{ij}$. Because these allowable values depend on $a_i + b_j$ the error $\varepsilon_{ij}$ cannot be independent of $a_i$ and $b_j$. It can, however, weight its allowable values in such a way that $E(\varepsilon_{ij} \mid a_i, b_j) = 0$ and $V(\varepsilon_{ij} \mid a_i, b_j) = \sigma^2_{E(i,j)}$. Therefore, a model with $\varepsilon_{ij}$ mutually independent from $(0, \sigma^2_{E(i,j)})$ conditionally on $a_1, \dots, a_R$ and $b_1, \dots, b_C$ is plausible. In such a model the $\varepsilon_{ij}$ are uncorrelated with $a_i$ and $b_j$.

The relaxation does not go quite as far as we might want. For example, if there is an upper bound on $X_{ij}$, then the largest possible $a_i$ and largest possible $b_j$ must sum to at most that bound, or else $\varepsilon_{ij}$ cannot have mean 0.

6.2. *Outer product models.* We might suppose that instead of the additive random effects model, that an outer product representation is more appropriate. Such SVD type models have become important in information retrieval [Deerwester et al. (1990)] and DNA microarray analysis [Alter, Brown and Botstein (2000)].

Under such a model we write

$$(11) \qquad X_{ij} = \mu + a_i + b_j + \sum_{\ell=1}^{L} \lambda_\ell u_{i\ell} v_{j\ell} + \varepsilon_{ij},$$

where the new pieces are scalar singular values $\lambda_\ell$ and the singular vectors with components $u_{i\ell}$ and $v_{j\ell}$. Ordinarily the singular vectors are fit to the data subject to a norm constraint. As a model for how the data might have arisen, we don't have to impose that constraint. We can make the pieces random with $u_{i\ell} \sim (0, \tau^2_{U(i,\ell)})$, and $v_{j\ell} \sim (0, \tau^2_{V(j,\ell)})$ independently of each other and the $a_i$ and $b_j$.

The model (11) is popular in crop science [Crossa and Cornelius (2002)] where the rows and columns correspond to genotypes and environments. Surprisingly, to modern readers that model (with $L = 1$) is about as old as the earliest ANOVA papers, going back to Fisher and Mackenzie (1923).

Writing

$$\eta_{ij} = \sum_{\ell=1}^{L} \lambda_\ell u_{i\ell} v_{j\ell} + \varepsilon_{ij},$$

we find that $\eta_{ij}$ are uncorrelated with each other and with $a_i$ and $b_j$. Uncorrelated errors $\eta_{ij}$ lead to the same variances as independent ones do. Therefore, we can subsume randomly generated outer products into the $\varepsilon_{ij}$ term of model (1). As a consequence, we can apply the pigeonhole bootstrap without knowing what the value of $L$ is, including the possibility that $L = 0$ might be the best description of the data.

A very early outer product model is the one degree of freedom for non-additivity of Tukey (1949). In this model

$$(12) \qquad X_{ij} = \mu + a_i + b_j + \lambda a_i b_j + \varepsilon_{ij}.$$

It differs from the additive plus outer product model in having the same variables appear in both places. Once again, we can subsume the outer product part into the error because $\lambda a_i b_j + \varepsilon_{ij}$ is uncorrelated with $a_i$, $b_j$ and $\lambda a_{i'} b_j + \varepsilon_{i'j}$ when $i \neq i'$, if all the $a_i$, $b_j$ and $\varepsilon_{ij}$ are independent.

6.3. *Bias.* It is a commonplace that biases affect overall levels much more than they affect comparisons. The EPA used to say about automobile efficiency that while "your mileage may vary" from what they report, reported differences between vehicles should be accurate. (Now they say "your mpg *will* vary.") In practice with the bootstrap, we would like to know whether a given statistic is performing with bias like the Tuesday score, or with much less bias like the Sunday versus Tuesday difference. For a scalar parameter we can compare the resampled values to the original one.

For our hypothetical bookseller wondering whether classical music lovers are more likely to purchase a Harry Potter book on a Tuesday, it now becomes clear that "more likely than what?" is an important consideration. A contrast with other days, other books or other customer types will be better determined than the absolute level.

It would be interesting to know whether the bias in the pigeonhole bootstrap tracks with the sampling bias in any reasonable generative model having random sample sizes. The mixed model (1) with fixed sample sizes does not allow for possibility of bias in sample means.

6.4. *Random observation patterns.* The analysis here is conditional on the values of $Z_{ij}$. The conditional and unconditional variances of $\hat{\mu}_x$ can be very different. When that happens the pigeonhole bootstrap will estimate the conditional variance which may differ greatly from the unconditional one, when $Z_{ij}$ is correlated with the response variable.

For the Netflix data, it is clear that there are strong dependencies between $Z_{ij}$ and $Y_{ij}$. Most people only rate movies they've seen, and those tend to be ones that they like or think they will like. A few people might be more likely to supply ratings for the movies they do not like, hoping to educate the algorithm about their tastes. Probably a few people spam the ratings to boost some movies and/or harm others. But on the whole, positive correlation is expected.

If we want to understand the values of $Y_{ij}$ for $ij$ pairs that were not observed, then an unconditional analysis accounting for varying $Z_{ij}$ is appropriate. If we want to predict $Y_{ij}$ ratings that will be made later, then conditioning is appropriate, because those future ratings will have similar, if not the same, selection bias.

Sometimes the unconditional problem is much more interesting. In the extreme, the unconditional analysis is essential when all we observe are the $Z_{ij}$ and we want to study co-ocurrence. But the conditional problem is interesting too. For example, the Netflix competition is only about predicting ratings that were actually made and then held out, not about predicting rating values that might have been made. So it is a conditional estimation problem. Similarly, in e-commerce, $Z_{ij} = 1$ may mean that customer $i$ saw an ad for product $j$, and the retailer studying what happens next is doing a conditional inference.

In some cases the conditional and unconditional variances can be expected to be close. Let $\mathcal{Z}$ represent $Z_{ij}$ for $i = 1, \ldots, R$ and $j = 1, \ldots, C$. Letting $\mathcal{Z}$ be random, we write

$$V(\hat{\mu}_x) = E(V_{\mathrm{RE}}(\hat{\mu}_x \mid \mathcal{Z})) + V(E_{\mathrm{RE}}(\hat{\mu}_x \mid \mathcal{Z})).$$

When $Z_{i\hat{j}} = 0$ describes a "missing at random" phenomenon, then $E_{\mathrm{RE}}(\hat{\mu}_x \mid \mathcal{Z}) = \mu$ has zero variance. Combining missing at random with the random effects model, we get

$$(13) \quad V(\hat{\mu}_x) = E\left(\frac{1}{N^2} \sum_{i=1}^{R} n_{i\bullet}^2 \sigma_{A(i)}^2 + \sum_{j=1}^{C} n_{\bullet j}^2 \sigma_{B(i)}^2 + \sum_{i=1}^{R} \sum_{j=1}^{C} Z_{ij} \sigma_{E(i,j)}^2\right),$$

which reduces to

$$(14) \qquad E\left(\frac{1}{N}(\nu_A \sigma_A^2 + \nu_B \sigma_B^2 + \sigma_E^2)\right)$$

for a homogenous random effects model. When the quantity within the expectations in (13) or in (14) is stable under sampling of $\mathcal{Z}$, then the conditional variance estimated by the pigeonhole bootstrap will be close to the unconditional one.

## APPENDIX: PROOF OF THEOREM 3

THEOREM 3. *The expected value under the random effects model of the pigeonhole variance is*

$$E_{\mathrm{RE}}(\hat{V}_{\mathrm{PB}}(\hat{\mu}_x^*)) = \frac{1}{N^2}\left[\sum_i \sigma_{A(i)}^2 \lambda_i^A + \sum_j \sigma_{B(j)}^2 \lambda_j^B + \sum_i \sum_j Z_{ij} \sigma_{E(i,j)}^2 \lambda_{i,j}^E\right],$$

*where*

$$\lambda_i^A = \left(1 - \frac{1}{C}\right) n_{i\bullet}^2 \left(1 - 2\frac{n_{i\bullet}}{N} + \frac{\nu_A}{N}\right) + \left(1 - \frac{1}{R}\right)\left(n_{i\bullet} - 2\mu_{i\bullet} n_{i\bullet} + \frac{\nu_B n_{i\bullet}^2}{N}\right)$$

$$+ n_{i\bullet}\left(1 - \frac{n_{i\bullet}}{N}\right)^2 + \frac{n_{i\bullet}^2}{N^2}(N - n_{i\bullet}),$$

$$\lambda_j^B = \left(1 - \frac{1}{R}\right)n_{\bullet j}^2\left(1 - 2\frac{n_{\bullet j}}{N} + \frac{\nu_B}{N}\right) + \left(1 - \frac{1}{C}\right)\left(n_{\bullet j} - 2\mu_{\bullet j}n_{\bullet j} + \frac{\nu_A n_{\bullet j}^2}{N}\right)$$

$$+ n_{\bullet j}\left(1 - \frac{n_{\bullet j}}{N}\right)^2 + \frac{n_{\bullet j}^2}{N^2}(N - n_{\bullet j})$$

and for $Z_{ij} = 1$,

$$\lambda_{i,j}^E = \left(1 - \frac{1}{C}\right)\left(1 - 2\frac{n_{i\bullet}}{N} + \frac{\nu_A}{N}\right)$$

$$+ \left(1 - \frac{1}{R}\right)\left(1 - 2\frac{n_{\bullet j}}{N} + \frac{\nu_B}{N}\right) + 1 - \frac{1}{N}.$$

PROOF. First,

$$n_{i\bullet}(\bar{X}_{i\bullet} - \hat{\mu}_x)$$

$$= \sum_j Z_{ij}X_{ij} - \frac{n_{i\bullet}}{N}\sum_{i'}\sum_j Z_{i'j}X_{i'j}$$

$$= \sum_{i'}\sum_j Z_{i'j}X_{i'j}\left(1_{i'=i} - \frac{n_{i\bullet}}{N}\right)$$

$$= \sum_{i'}\sum_j Z_{i'j}(\mu + a_{i'} + b_j + \varepsilon_{i'j})\left(1_{i'=i} - \frac{n_{i\bullet}}{N}\right)$$

$$= \sum_{i'} a_{i'}\sum_j Z_{i'j}\left(1_{i'=i} - \frac{n_{i\bullet}}{N}\right) + \sum_j b_j\sum_{i'} Z_{i'j}\left(1_{i'=i} - \frac{n_{i\bullet}}{N}\right)$$

$$+ \sum_{i'}\sum_j \varepsilon_{i'j}Z_{i'j}\left(1_{i'=i} - \frac{n_{i\bullet}}{N}\right)$$

$$= \sum_{i'} a_{i'}n_{i'\bullet}\left(1_{i'=i} - \frac{n_{i\bullet}}{N}\right) + \sum_j b_j\left(Z_{ij} - \frac{n_{\bullet j}n_{i\bullet}}{N}\right)$$

$$+ \sum_{i'}\sum_j \varepsilon_{i'j}Z_{i'j}\left(1_{i'=i} - \frac{n_{i\bullet}}{N}\right).$$

Therefore,

$$E_{\mathrm{RE}}\left(\sum_i n_{i\bullet}^2(\bar{X}_{i\bullet} - \hat{\mu}_x)^2\right)$$

$$= \sum_i\sum_{i'} \sigma_{A(i')}^2 n_{i'\bullet}^2\left(1_{i'=i} - 2 \cdot 1_{i'=i}\frac{n_{i\bullet}}{N} + \frac{n_{i\bullet}^2}{N^2}\right)$$

$$+ \sum_i \sum_j \sigma_{B(j)}^2 \left( Z_{ij} - 2 Z_{ij} \frac{n_{i\bullet} n_{\bullet j}}{N} + \frac{n_{i\bullet}^2 n_{\bullet j}^2}{N^2} \right)$$

$$+ \sum_i \sum_{i'} \sum_j \sigma_{E(i',j)}^2 Z_{i'j} \left( 1_{i'=i} - 2 \cdot 1_{i'=i} \frac{n_{i\bullet}}{N} + \frac{n_{i\bullet}^2}{N^2} \right)$$

$$= \sum_{i'} \sigma_{A(i')}^2 n_{i'\bullet}^2 \left( 1 - 2 \frac{n_{i'\bullet}}{N} + \frac{\nu_A}{N} \right)$$

$$+ \sum_j \sigma_{B(j)}^2 \left( n_{\bullet j} - 2 \mu_{\bullet j} n_{\bullet j} + \frac{\nu_A n_{\bullet j}^2}{N} \right)$$

$$+ \sum_{i'} \sum_j \sigma_{E(i',j)}^2 Z_{i'j} \left( 1 - 2 \frac{n_{i'\bullet}}{N} + \frac{\nu_A}{N} \right),$$

and the analogous expression holds for $E_{\mathrm{RE}}(\sum_j n_{\bullet j}^2 (\bar{X}_{\bullet j} - \hat{\mu}_x)^2)$. Next, for cases with $Z_{ij} = 1$, $E_{\mathrm{RE}}((X_{ij} - \hat{\mu}_x)^2)$ equals

$$E_{\mathrm{RE}} \left( \left[ a_i + b_j + \varepsilon_{ij} - \frac{1}{N} \sum_{i'} \sum_{j'} Z_{i'j'} (a_{i'} + b_{j'} + \varepsilon_{i'j'}) \right]^2 \right)$$

$$= \sigma_{A(i)}^2 (1 - n_{i\bullet}/N)^2 + \sigma_{B(j)}^2 (1 - n_{\bullet j}/N)^2 + \sigma_{E(i,j)}^2 (1 - 1/N)^2$$

$$+ \sum_{i' \neq i} \sigma_{A(i')}^2 \frac{n_{i'\bullet}^2}{N^2} + \sum_{j' \neq j} \sigma_{B(j')}^2 \frac{n_{\bullet j'}^2}{N^2} + \sum_{i'} \sum_{j'} (1 - 1_{i=i'} 1_{j=j'}) \sigma_{E(i',j')}^2 \frac{Z_{i'j'}}{N^2}.$$

Thus, $E_{\mathrm{RE}}(\sum_i \sum_j Z_{ij} (X_{ij} - \hat{\mu}_x)^2)$ equals

$$\sum_i \sigma_{A(i)}^2 \left[ n_{i\bullet} \left( 1 - \frac{n_{i\bullet}}{N} \right)^2 + \frac{n_{i\bullet}^2}{N^2} (N - n_{i\bullet}) \right]$$

$$+ \sum_j \sigma_{B(j)}^2 \left[ n_{\bullet j} \left( 1 - \frac{n_{\bullet j}}{N} \right)^2 + \frac{n_{\bullet j}^2}{N^2} (N - n_{\bullet j}) \right]$$

$$+ \sum_i \sum_j Z_{ij} \sigma_{E(i,j)}^2 \left[ \left( 1 - \frac{1}{N} \right)^2 + \frac{N-1}{N^2} \right].$$

Putting the pieces together and applying Theorem 2, $E_{\mathrm{RE}}(\hat{V}_{\mathrm{PB}}(\hat{\mu}_x^*))$ equals

$$\frac{1}{N^2} \left[ \sum_i \sigma_{A(i)}^2 \lambda_i^A + \sum_j \sigma_{B(j)}^2 \lambda_j^B + \sum_i \sum_j Z_{ij} \sigma_{E(i,j)}^2 \lambda_{i,j}^E \right],$$

where

$$\lambda_i^A = \left( 1 - \frac{1}{C} \right) n_{i\bullet}^2 \left( 1 - 2 \frac{n_{i\bullet}}{N} + \frac{\nu_A}{N} \right)$$

$$+ \left(1 - \frac{1}{R}\right)\left(n_{i\bullet} - 2\mu_{i\bullet}n_{i\bullet} + \frac{\nu_B n_{i\bullet}^2}{N}\right)$$

$$+ n_{i\bullet}\left(1 - \frac{n_{i\bullet}}{N}\right)^2 + \frac{n_{i\bullet}^2}{N^2}(N - n_{i\bullet}),$$

$$\lambda_j^B = \left(1 - \frac{1}{R}\right)n_{\bullet j}^2\left(1 - 2\frac{n_{\bullet j}}{N} + \frac{\nu_B}{N}\right)$$

$$+ \left(1 - \frac{1}{C}\right)\left(n_{\bullet j} - 2\mu_{\bullet j}n_{\bullet j} + \frac{\nu_A n_{\bullet j}^2}{N}\right)$$

$$+ n_{\bullet j}\left(1 - \frac{n_{\bullet j}}{N}\right)^2 + \frac{n_{\bullet j}^2}{N^2}(N - n_{\bullet j}) \quad \text{and for } Z_{ij} = 1,$$

$$\lambda_{i,j}^E = \left(1 - \frac{1}{C}\right)\left(1 - 2\frac{n_{i\bullet}}{N} + \frac{\nu_A}{N}\right)$$

$$+ \left(1 - \frac{1}{R}\right)\left(1 - 2\frac{n_{\bullet j}}{N} + \frac{\nu_B}{N}\right) + 1 - \frac{1}{N}. \qquad \square$$

**Acknowledgments.** I thank Netflix for making available their movie ratings data set. I thank Peter McCullagh and Patrick Perry for helpful discussions. Thanks to the editor for encouraging a deeper analysis of the Tuesday effect.

## REFERENCES

ALTER, O., BROWN, P. O. and BOTSTEIN, D. (2000). Singular value decomposition for genome-wide expression data processing and analysis. *PNAS* **97** 10101–10106.

COCHRAN, W. G. (1977). *Sampling Techniques*, 3rd ed. Wiley, New York. MR0474575

CORNFIELD, J. and TUKEY, J. W. (1956). Average values of mean squares in factorials. *Ann. Math. Statist.* **27** 907–949. MR0087282

CROSSA, J. and CORNELIUS, P. L. (2002). Linear–bilinear models for the analysis of genotype-environment interaction data. In *Quantitative Genetics, Genomics and Plant Breeding in the 21st Century, an International Symposium* (M. S. Kang, ed.) 305–322. CAB International, Wallingford UK.

DEERWESTER, S., DUMAIS, S., FURNAS, G. W., LANDAUER, T. K. and HARSHMAN, R. (1990). Indexing by latent semantic analysis. *J. Soc. Inform. Sci.* **41** 391–407.

DHILLON, I. S. (2001). Co-clustering documents and words using bipartite spectral graph partitioning. In *Proceedings of the Seventh ACM SIGKDD International Conference on Knowledge Discovery and Data Mining (KDD)*.

FISHER, R. A. and MACKENZIE, W. A. (1923). The manurial response of different potato varieties. *J. Agricultural Science* **XIII** 311–320.

MCCULLAGH, P. (2000). Resampling and exchangeable arrays. *Bernoulli* **6** 285–301. MR1748722

SEARLE, S. R., CASELLA, G. and MCCULLOCH, C. E. (1992). *Variance Components*. Wiley, New York. MR1190470

Tukey, J. W. (1949). One degree of freedom for non-additivity. *Biometrics* **5** 232–242.

Department of Statistics
Stanford University
Stanford, California 94305-4065
USA
E-mail: owen@stat.stanford.edu